\documentclass[useAMS,usenatbib,letterpaper]{mn2e}

\usepackage{graphicx}
\usepackage{amsmath}

\newcommand\apjl{ApJ}%
\newcommand\apjs{ApJS}%
\newcommand\aap{A\&A}%
\def\ESO{1}
\def\Carnegie{2}
\def\UPenn{3}
\def\CfA{4}
\def\KICPChicago{5}
\def\EFIChicago{6}
\def\Arizona{7}
\def\Cavendish{8}
\def\Miss{9}
\def\PhysicsUChicago{10}
\def\AAUChicago{11}
\def\ANL{12}
\def\Cambridge{13}
\def\Dal{14}
\def\Davis{15}
\def\UFlorida{16}
\def\UCL{17}
\def\McGill{18}
\def\Berkeley{19}
\def\UCLA{20}
\def\ATNF{21}
\def\Harvard{22}
\def\Caltech{23}
\def\MPIfR{24}

\title[CO two bright, lensed star-forming galaxies at $z\sim2.7$]{Large gas reservoirs and free-free emission in two lensed star-forming galaxies at $z=2.7$}
\author[M. Aravena et al.]
{\parbox{\textwidth}{M.~Aravena$^{\ESO}$\thanks{E-mail: maravena@eso.org}, 
 E.~J.~Murphy$^{\Carnegie}$,
 J.~E.~Aguirre$^{\UPenn}$,
 M.~L.~N.~Ashby$^{\CfA}$,
B.~A.~Benson$^{\KICPChicago,\EFIChicago}$, 
 M.~Bothwell$^{\Arizona,\Cavendish}$,
M.~Brodwin$^{\Miss}$,
J.~E.~Carlstrom$^{\KICPChicago,\PhysicsUChicago,\EFIChicago,\AAUChicago,\ANL}$, 
 S.~C.~Chapman$^{\Dal,\Cambridge}$,
T.~M.~Crawford$^{\KICPChicago,\AAUChicago}$, 
 C.~de~Breuck$^{\ESO}$,
 C.~D.~Fassnacht$^{\Davis}$,
A.~H.~Gonzalez$^{\UFlorida}$, 
 T.~R.~Greve$^{\UCL}$,
 B.~Gullberg$^{\ESO}$,	
 Y.~Hezaveh$^{\McGill}$,
G.~P.~Holder$^{\McGill}$, 
W.~L.~Holzapfel$^{\Berkeley}$, 
R.~Keisler$^{\KICPChicago,\PhysicsUChicago}$, 
 M.~Malkan$^{\UCLA}$,
 D.~P.~Marrone$^{\Arizona}$,   
 V.~McIntyre$^{\ATNF}$,
C.~L.~Reichardt$^{\Berkeley}$, 
 K.~Sharon$^{\KICPChicago,\AAUChicago}$,
 J.~S.~Spilker$^{\Arizona}$,
B.~Stalder$^{\CfA}$, 
A.~A.~Stark$^{\CfA}$, 
J.~D.~Vieira$^{\Caltech}$,
 A.~Wei\ss$^{\MPIfR}$
}\vspace{0.5cm}\\
\parbox{\textwidth}{
$^{\ESO}$ European Southern Observatory, , Alonso de Cordova 3107, Casilla 19001 Vitacura Santiago, Chile.\\
$^{\Carnegie}$ Observatories of the Carnegie Institution for Science, 813 Santa Barbara Street, Pasadena, CA 91101, USA\\
$^{\UPenn}$ University of Pennsylvania, 209 South 33rd Street, Philadelphia, PA 19104, USA\\
$^{\CfA}$ Harvard-Smithsonian Center for Astrophysics, 60 Garden Street, Cambridge, MA 02138, USA\\
$^{\KICPChicago}$ Kavli Institute for Cosmological Physics, University of Chicago, 5640 South Ellis Avenue, Chicago, IL 60637, USA\\
$^{\EFIChicago}$ Enrico Fermi Institute, University of Chicago, 5640 South Ellis Avenue, Chicago, IL 60637, USA\\
$^{\Arizona}$ Steward Observatory, University of Arizona, 933 North Cherry Avenue, Tucson, AZ 85721, USA\\
$^{\Cavendish}$ Cavendish Laboratory, University of Cambridge, 19 J.J. Thomson Avenue, Cambridge, CB3 0HE, UK\\
$^{\Miss}$ Department of Physics and Astronomy, University of Missouri, 5110 Rockhill Road, Kansas City, MO 64110, USA\\
$^{\PhysicsUChicago}$ Department of Physics, University of Chicago, 5640 South Ellis Avenue, Chicago, IL 60637, USA\\
$^{\AAUChicago}$ Department of Astronomy and Astrophysics, University of Chicago, 5640 South Ellis Avenue, Chicago, IL 60637, USA\\
$^{\ANL}$ Argonne National Laboratory, 9700 S. Cass Avenue, Argonne, IL, USA 60439, USA\\
$^{\Cambridge}$ Institute of Astronomy, University of Cambridge, Madingley Road, Cambridge CB3 0HA, UK\\
$^{\Dal}$ Dalhousie University, Halifax, Nova Scotia, Canada\\
$^{\Davis}$ Department of Physics,  University of California, One Shields Avenue, Davis, CA 95616, USA\\
$^{\UFlorida}$ Department of Astronomy, University of Florida, Gainesville, FL 32611, USA\\
$^{\UCL}$ Department of Physics and Astronomy, University College London, Gower Street, London WC1E 6BT, UK\\
$^{\McGill}$ Department of Physics, McGill University, 3600 Rue University, Montreal, Quebec H3A 2T8, Canada\\
$^{\Berkeley}$ Department of Physics, University of California, Berkeley, CA 94720, USA\\
$^{\UCLA}$ Department of Physics and Astronomy, University of California, Los Angeles, CA 90095-1547, USA\\
$^{\ATNF}$ Australia Telescope National Facility, CSIRO, Epping, NSW 1710, Australia\\
$^{\Harvard}$ Department of Physics, Harvard University, 17 Oxford Street, Cambridge, MA 02138, USA\\
$^{\Caltech}$ California Institute of Technology, 1200 E. California Blvd., Pasadena, CA 91125, USA\\
$^{\MPIfR}$ Max-Planck-Institut f\"{u}r Radioastronomie, Auf dem H\"{u}gel 69 D-53121 Bonn, Germany
}}

\begin{document}


\maketitle
\label{firstpage}
\begin{abstract}
We report the detection of CO(1-0) line emission in the bright, lensed star-forming galaxies  SPT-S\,233227-5358.5 ($z=2.73$) and SPT-S\,053816-5030.8 ($z=2.78$), using the Australia Telescope Compact Array (ATCA). Both galaxies were discovered in a large-area millimeter survey with the South Pole Telescope (SPT) and found to be gravitationally lensed by intervening structures. The measured CO  intensities imply galaxies with molecular gas masses of $(3.2\pm0.5)\times10^{10}(\mu/15)^{-1}(X_\mathrm{CO}/0.8)\,M_\odot$ and $(1.7\pm0.3)\times10^{10}(\mu/20)^{-1}(X_\mathrm{CO}/0.8)\,M_\odot$, and gas depletion timescales of $4.9\times10^7(X_\mathrm{CO}/0.8)$ yr and $2.6\times10^7(X_\mathrm{CO}/0.8)$ yr, respectively, where $\mu$ corresponds to the lens magnification and $X_\mathrm{CO}$ is the CO luminosity to gas mass conversion factor. In the case of SPT-S\,053816-5030.8, we also obtained significant detections of the rest-frame 115.7 and 132.4\,GHz radio continuum. Based on the radio to infrared spectral energy distribution and an assumed synchrotron spectral index, we find that $42\pm10\%$ and $55\pm13\%$ of the flux at rest-frame 115.7 and 132.4\,GHz arise from free-free emission. We find a radio-derived intrinsic star formation rate (SFR) of $470\pm170~M_\odot$ yr$^{-1}$, consistent within the uncertainties with the infrared estimate. Based on the morphology of this object in the source plane, the derived gas mass and the possible flattening of the radio spectral index towards low frequencies, we argue that SPT-S\,053816-5030.8 exhibits properties compatible with a scaled-up local ultra-luminous infrared galaxy. 
\end{abstract}


\section{Introduction}
Submillimeter/millimeter blank field surveys have uncovered a significant population of dusty star-forming galaxies (DSFGs) at cosmological distances. These galaxies are found to harbor intense star formation activity, with typical SFRs $>500\ M_\odot$ yr$^{-1}$, and to be generally located at $z>1$ \citep{blain02,chapman05,lagache05,vieira13,weiss13}.  Their observational properties suggest they are the likely progenitors of local massive early type galaxies, and they appear to trace large galaxy overdensities at high redshifts \citep{brodwin08,viero09,daddi09a, aravena10, capak11,amblard11}.

The interstellar medium (ISM) plays a critical role in shaping the overall evolution of galaxies as it constitutes the primary reservoir of material for star formation. The continuous interaction between stars and the ISM governs the overall and chemical composition of galaxies. The molecular gas in the ISM is mostly composed of H$_2$ and He, with a minor fraction ($\sim1\%$) of other molecules and dust. However, H$_2$ is difficult to detect in the cold molecular ISM. The CO molecule, particularly in its lowest rotational transition $J=1-0$, represents the best tracer of H$_2$, specifically its mass and spatial distribution \citep[e.g., ][]{omont07}.

Large efforts have been devoted to investigating the physical properties of DSFGs. However, due to their faintness and large cosmological distances, only a handful have been targeted in a sufficiently large number of molecular gas emission lines (e.g., CO) to allow for a full characterization of their ISM. Follow-up CO studies have been focused on the observation of $J\geq3$ CO transitions. However, CO(1-0) observations in a few unlensed DSFGs indicate that the gas masses are $2\times$ larger and up to $3\times$ more spatially extended than expected based on $J\geq3$ CO transitions, heavily altering our understanding of these systems \citep[e.g.,][]{papadopoulos02,greve03,harris10,ivison11,riechers11b}. This highlights the need for low-J CO observations.

Strongly lensed systems, including IRAS F10214+4722, the Cloverleaf, the Cosmic Eyelash and SMM 02399-0136, have proven to be a boon for studies of far-infrared (IR) and millimeter line studies at high redshift \citep[e.g., ][]{solomon05,brown91,rowanrobinson91,brown92,barvainis92,ivison98,frayer98,swinbank10}. The recent discovery of a population of gravitationally lensed, highly magnified, star-forming galaxies in the $>100$ square degree millimeter and far-IR surveys carried out by the South Pole Telescope \citep[SPT; ][]{carlstrom11} and the {\it Herschel Space Observatory} \citep{pilbratt10} has provided a remarkable opportunity to study in great detail the process of star-formation and conditions of the ISM in this cosmologically important population of galaxies \citep{vieira10,negrello10}.

We have started a systematic follow-up campaign with the Australia Telescope Compact Array (ATCA) to study the molecular gas content in a sample of gravitationally lensed star-forming galaxies through the observation of $J\leq2$ CO emission. These galaxies were found in the 1.4-3~mm wavelength sky maps produced by the SPT Sunyaev-Zel'dovich effect survey and have accurate spectroscopic redshifts \citep{greve12}. In this work, we report pilot observations of the CO(1-0) line emission in two such sources. Hereafter, we adopt a $\Lambda$CDM cosmology with $H_0=71$ km s$^{-1}$ Mpc$^{-1}$, $\Omega_\mathrm{M}=0.27$, and $\Omega_\Lambda=0.73$.

\begin{figure*}
\vspace{2mm}
\centering
\includegraphics[scale=0.5]{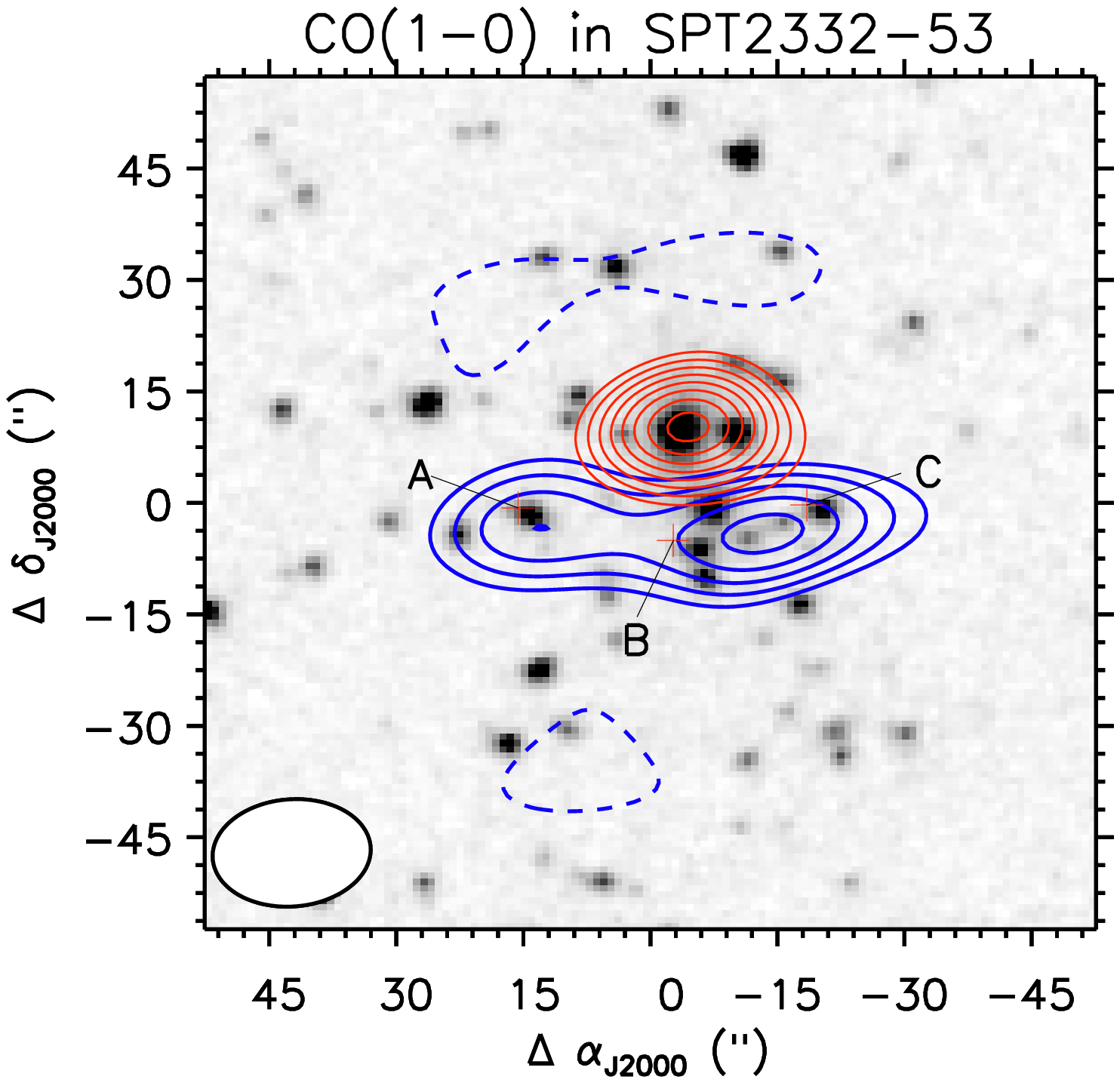}\hspace{5mm}
\includegraphics[scale=0.5]{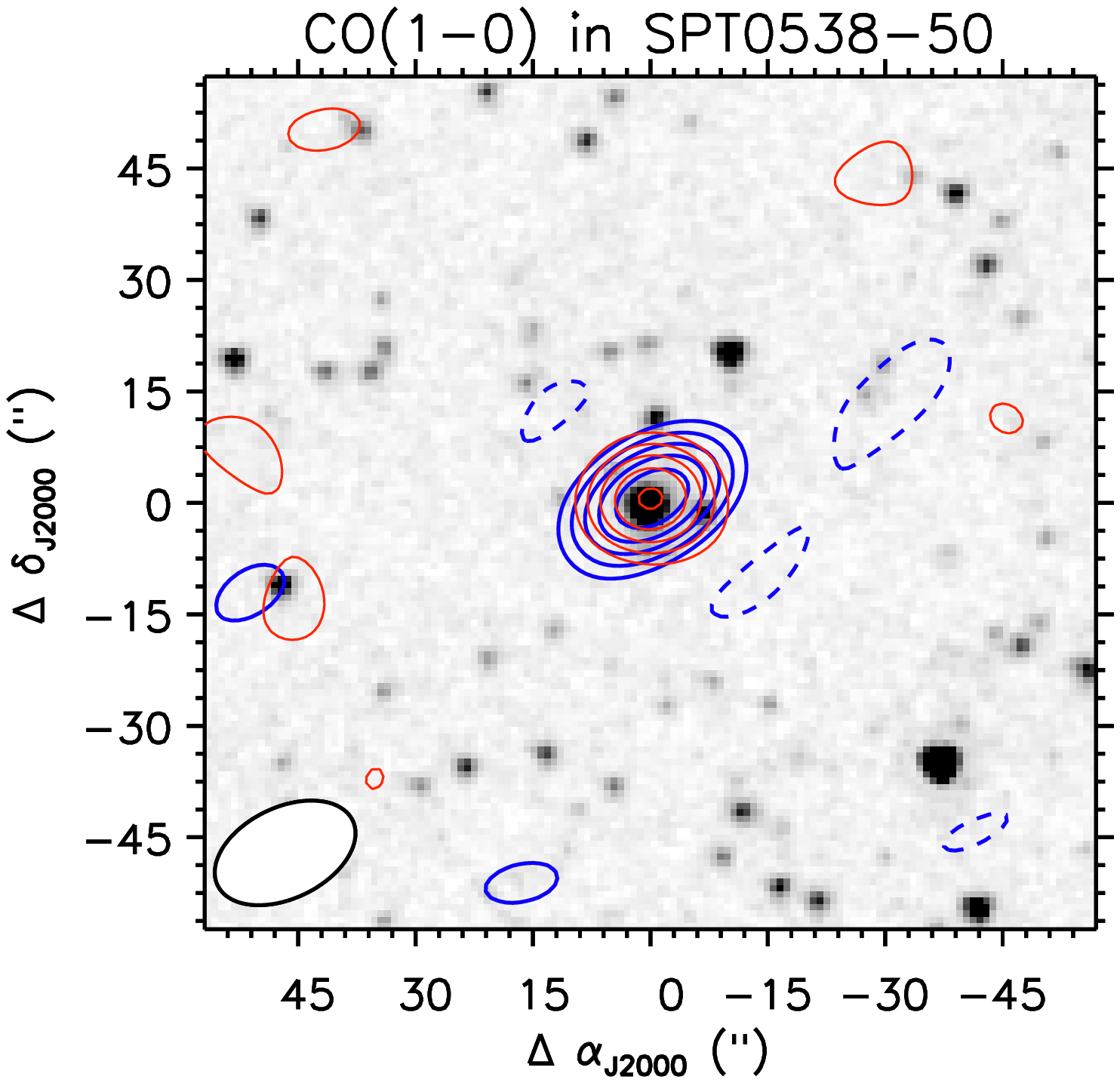}
\vspace{2mm}
\caption{CO(1-0) and 35 GHz continuum detections toward SPT\,2332-53 and SPT\,0538-50. The {\it Spitzer} IRAC 4.5$\mu$m image is shown in greyscale with blue contours overlaid representing the CO(1-0) line emission (dirty image) averaged over 350 km s$^{-1}$ and 220 km s$^{-1}$, respectively. These contours are in steps of $\pm1\sigma$, starting at $\pm2\sigma$, with $\sigma=0.32$ mJy beam$^{-1}$ and $\sigma=0.44$ mJy beam$^{-1}$, respectively. Red contours represent the 35 GHz continuum emission (dirty image), and are shown in steps of $+1\sigma$, starting at $+2\sigma$, with $\sigma=31\ \mu$Jy beam$^{-1}$ and $\sigma=21\ \mu$Jy beam$^{-1}$, respectively. For SPT\ 2332-53, the location of the SABOCA sources are shown as red plus symbols, following the notation from \citet{greve12}. The axes show the offset, in arcseconds, with respect to the phase tracking center. \label{fig:co_image}}
\end{figure*}

\section{Observations}

\subsection{SPT sources}
The two targets of this work, SPT-S\,233227$-$5358.5 and SPT-S\,053816$-$5030.8 (hereafter SPT\,2332-53 and SPT\,0538-50), were the first dusty-spectrum SPT sources \citep{vieira10} to have spectroscopic redshift measurements. In both cases, a redshift determination was obtained through the identification of CO and [CI] line emission in wide-band submillimeter spectra obtained with the Z-spec instrument on the Atacama Pathfinder Experiment (APEX) 12m telescope, and confirmed with optical spectroscopy obtained with the Very Large Telescope (VLT). The redshifts derived for the lensed objects are $z=2.738$ and $z=2.783$ for SPT\,2332-53 and SPT\,0538-50, respectively \citep{greve12}.  Both galaxies are associated with faint {\it Spitzer} Infrared Array Camera (IRAC) 4.5$\mu$m sources (Fig.~1), and are detected with the APEX Large Bolometer Camera (LABOCA) and the Submillimeter APEX Bolometer Camera (SABOCA) at 870$\mu$m and $350\mu$m, respectively \citep[see][]{greve12}. 

As shown by the 850$\mu$m and $350\mu$m images \citep[Fig. \ref{fig:co_image} in ][]{greve12}, the submillimeter emission in SPT\,2332-53 is resolved into six major components, three of which are detected at 350$\mu$m  (A, B and C; Fig. \ref{fig:co_image}). A massive foreground galaxy cluster at $z =0.403$ is identified as the gravitational lens for this source \citep{greve12}. A full analysis of the lensing structure of this object will be presented in Vieira et al. (in preparation). That work leads to a magnification factor in the range $\mu=10-20$, and thus we adopt $\mu=15\pm5$ throughout. 

SPT\ 0538-50 is unresolved in the LABOCA and SABOCA images. Recent ALMA 850 $\mu$m sub-arcsecond imaging of this source shows an Einstein ring around a foreground galaxy at $z=0.443$, confirming the gravitational lensed nature of this source \citep{hezaveh13}. A detailed lens model gives a magnification $\mu=20\pm4$ \citep{hezaveh13}. Unless otherwise stated, the quoted uncertainties in the magnification factors for both sources are not propagated to the observed quantities throughout.

\citet{greve12} find observed IR luminosities $L_\mathrm{IR}(8-1000\mu\mathrm{m})=(6.5\pm1.6)\times10^{13}\ L_\odot$ for both sources. However, recent {\it Herschel} far-IR data leads to an improved observed IR luminosity of $(8.8\pm1.3)\times10^{13}\ L_\odot$ for SPT\ 0538-50 \citep{hezaveh13}. The revised value for the IR luminosity of SPT2332-53 is well within uncertainties with the estimate from \citet{greve12}, and we use this value throughout. From this, we adopt intrinsic IR luminosities of $(4.3\pm1.1)\times10^{12} (\mu/15)^{-1}\ L_\odot$ and $(4.4\pm0.65)\times10^{12} (\mu/20)^{-1}\ L_\odot$ for SPT\ 2332-53 and SPT\ 0538-50, respectively. The estimated total IR luminosities are derived from fitting a single-temperature grey-body model to the data at $\lambda_\mathrm{rest}>50\ \mu$m, and thus do not account for a possible hot dust component \citep[see][for details]{greve12}. While such hot dust component would increase the actual IR luminosities of our targets, such emission would likely arise from AGN and thus would not strictly trace the star formation in the host galaxies, being irrelevant for the analysis below.

\subsection{ATCA CO data}

We used ATCA in the H75 compact-array configuration to observe the redshifted CO(1-0) emission line ($\nu_\mathrm{rest}=115.271$ GHz) in the two star-forming galaxies SPT\ 2332-53 and SPT\ 0538-50. The observations were performed between UT 2012 July 27-30 under good weather conditions with 5 working antennas, and the sources were observed for $\sim$10h each. 

We used the Compact Array Broadband Backend (CABB) configured in the 1M-0.5k mode, which leads to a bandwidth of 2 GHz per correlator window with 1 MHz per channel of spectral resolution. In both cases, one of the windows was tuned to 35000 MHz to obtain a measurement of the continuum emission. For SPT\ 2332-53 the other window was tuned to 31412 MHz, and for SPT\ 0538-50 this was tuned to 30600 MHz. The phase tracking centers were $\alpha_\mathrm{2000}=23^h32^m27.8^s$, $\delta_\mathrm{2000}=-53^\circ58'37''$, and $\alpha_\mathrm{2000}=05^h38^m16.8^s$, $\delta_\mathrm{2000}=-50^\circ30'52''$, respectively, based on the prior positions obtained with LABOCA at 870$\mu$m. 

The bright quasars 2355-534 ($S_\mathrm{35GHz}\sim1.8$ Jy) and 0537-441 ($S_\mathrm{35GHz}\sim6.2$ Jy) were observed every 6-7 min for gain calibration, and every $\sim1$h for pointing calibration.  The sources 1921-293 ($\sim14$ Jy) and 0537-441 were used for bandpass calibration, and Uranus and 1934-638 were used as amplitude calibrators. The software package {\it Miriad} \citep{sault95} and the Common Astronomy Software Applications \citep[CASA;][]{mcmullin07} were used for editing, calibration, and imaging. Absolute flux calibration is estimated to be accurate to within $15\%$.

Flagging of visibilities was performed for the case of SPT\ 0538-50, where some channels fell close to the edge of the Q-band short frequency limit at 30 GHz. The visibilities were inverted using natural weighting yielding synthesized beam sizes for SPT\ 2332-53 and SPT\ 0538-50 of $21.8''\times16.8''$ and $22.2''\times13.9''$ (at $\sim31$ GHz), and $19.2''\times14.8''$ and $18.0''\times14.2''$ (at 35 GHz), respectively. The final data results in an rms of 0.70 mJy beam$^{-1}$ per 4 MHz (39 km s$^{-1}$) channel at $\sim31.4$ GHz and 0.60 mJy beam$^{-1}$ per 2 MHz (19.7 km s$^{-1}$) channel at $\sim30.6$ GHz for SPT\,2332-53 and SPT\ 0538-50. We reach continuum rms levels of $40\ \mu$Jy beam$^{-1}$ and $35\ \mu$Jy beam$^{-1}$ at 31.4 and 35 GHz, respectively, for SPT\,2332-53, and $35\ \mu$Jy beam$^{-1}$ and $20\ \mu$Jy beam$^{-1}$ at 30.6 and 35 GHz, respectively, for SPT\,0538-50.

\subsection{ATCA low frequency data}
ATCA observations at 2.2, 5.5, and 9.0\,GHz of SPT\ 2332-53 and SPT\ 0538-50 were made between UT 2012 January 23-27 using the CABB in the 1M-0.5k mode (see above).  The observations were performed in the most extended ATCA configuration, 6A, with 6 working antennas. The data were edited, calibrated and imaged using the {\it Miriad} package. Data affected by known radio frequency interference (RFI) or with bad visibilility ranges were flagged accordingly. We estimate an absolute calibration uncertainty of $\sim5\%$ at 2.2 and 5.5\,GHz, and 10\% at 9.0\,GHz.  We inverted the visibilities using natural weighting leading to beam sizes of $\sim7.7''\times5.2''$, $3.2''\times2.1''$ and $2.0''\times1.3''$ at 2.2, 5.5 and 9.0 GHz, respectively. 
For SPT\ 2332-53, we reach rms noise values  of 200, 500, and 100\,$\mu$Jy beam$^{-1}$ at 2.2, 5.5, and 9.0\,GHz, respectively.  For SPT\ 0538-50, we reach rms noise values of 200, 100, and 75\,$\mu$Jy beam$^{-1}$ at each frequency, respectively.

\section{Results}

\begin{table*}
\centering
\caption{Observed CO properties\label{table:properties}}
\begin{tabular}{lcccccccc}
\hline
Source  & $z_\mathrm{CO}\ ^a$ & $S_\mathrm{CO}\ ^b$ & $v_\mathrm{FWHM}\ ^b$ & $I_{\mathrm{CO}\ 1-0}\ ^c$ & $L_\mathrm{CO}'\ ^d$  & $S_{31 \mathrm{GHz}}\ ^e$ & $S_{35 \mathrm{GHz}}\ ^e$ \\
           &                  & (mJy) & km s$^{-1}$  & (Jy km s$^{-1}$)       & ($\times10^{10}\ (\mu/\mu_0)^{-1}$ K km s$^{-1}$ pc$^2$)  & ($\mu$Jy) & ($\mu$Jy)   \\
        \hline\hline
SPT\ 2332-53$^\dagger$     & $2.7256(2)$ & $4.66\pm0.51$  & $342\pm42$ &  $1.70\pm0.25$  & $3.94\pm0.58$ & $<120$  &  $<120$    \\
SPT\ 0538-50blue$^\ddagger$   & $2.7854(1)$ & $3.81\pm0.36$  & $210\pm34$ &  $0.85\pm0.08$                  & $1.49\pm0.14$                  & $\ldots$ &  $\ldots$   \\
SPT\ 0538-50red$^\ddagger$   & $2.7897(6)$ & $1.18\pm0.27$  & $281\pm80$ &  $0.35\pm0.18$                  & $0.61\pm0.31$                   & $\ldots$ &  $\ldots$   \\
SPT\ 0538-50     & $2.7855(1)$  & $\ldots$               & $\ldots$         &  $1.20\pm0.20$   & $2.16\pm0.36$    &  $171\pm35$ & $130\pm20$ \\
\hline
\end{tabular}\\
\begin{flushleft}
\noindent $^\dagger$ Integrated values over all CO components are listed. \\
\noindent $^\ddagger$ Here, SPT\ 0538-50blue and SPT~0538-50red are the brighter and fainter CO peaks, respectively.\\
\noindent $^a$ CO redshift. For SPT\ 0538-50, the redshift is computed as the weighted average between the redshifts of SPT\ 0538-50blue and red.\\
\noindent $^b$ CO line peak flux and line full-width half maximum (FWHM) velocity. \\
\noindent $^c$ Integrated CO line intensity ($I_\mathrm{CO}=\int S_\mathrm{CO} dv$).\\
\noindent $^d$ CO line luminosity. The magnification factor $\mu_0$ is assumed to be 15 and 20 for SPT\ 2332-53 and SPT\ 0538-50, respectively. \\
\noindent $^e$ Continuum fluxes at $\sim31$ GHz and $35$ GHz. 
\end{flushleft}
\end{table*}

\begin{figure*}
\vspace{2mm}
\centering
\includegraphics[scale=0.5]{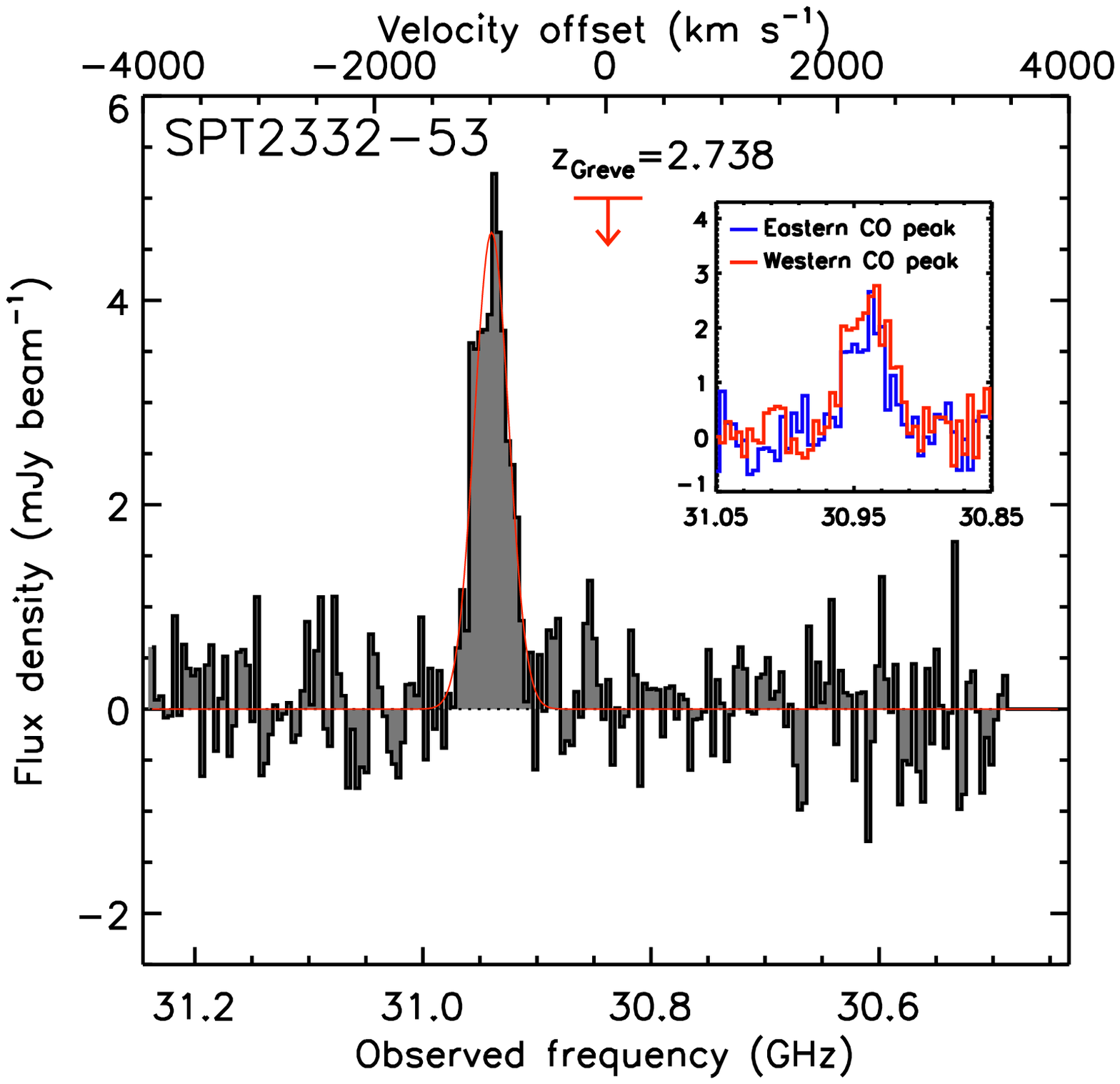}\hspace{3mm}
\includegraphics[scale=0.5]{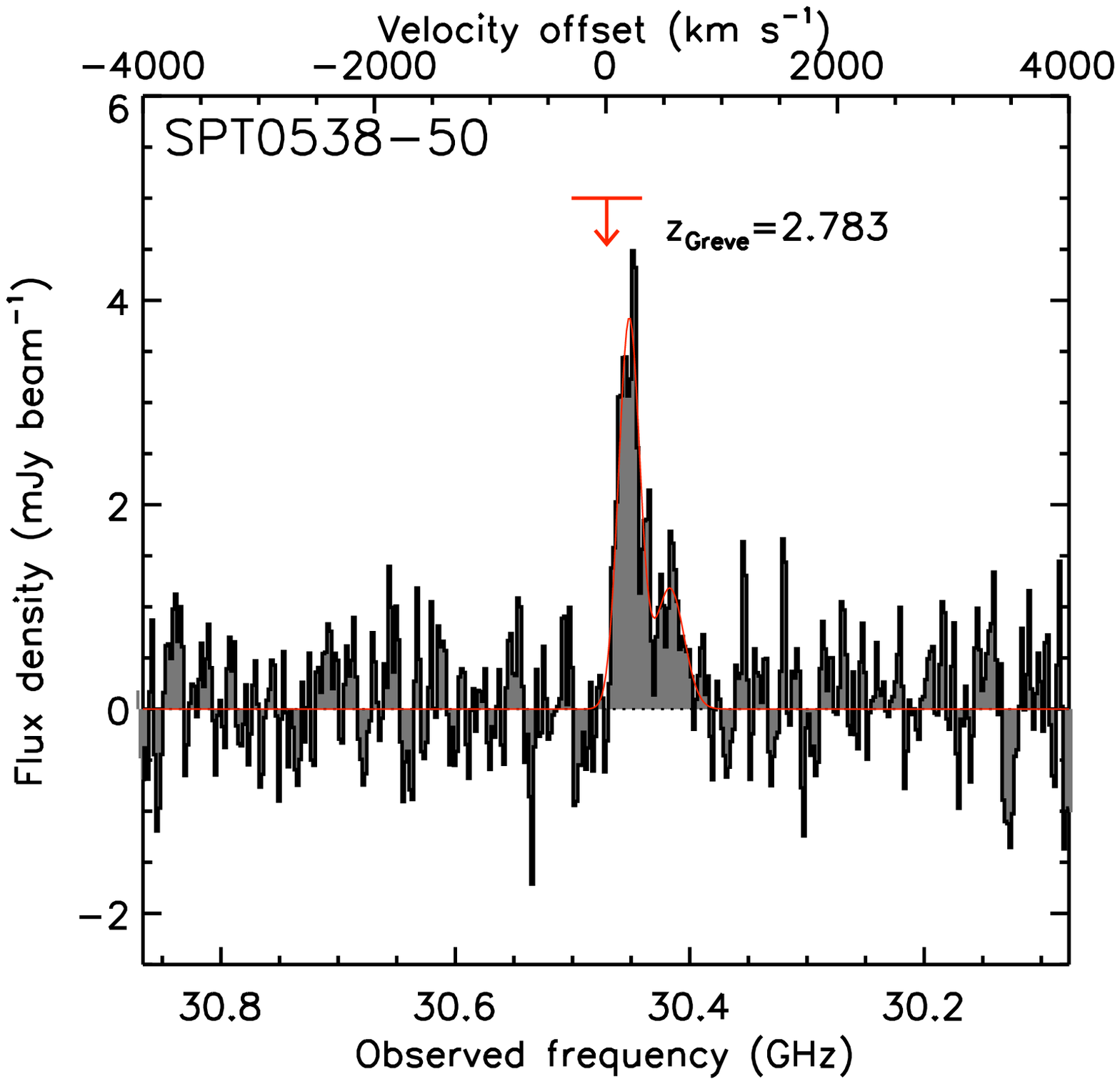}
\vspace{2mm}
\caption{ATCA spectra of the CO(1-0) line emission toward SPT\ 2332-53 and SPT\ 0538-50. Their spectra are shown at 39 km s$^{-1}$ (4 MHz) and 20 km s$^{-1}$ (2 MHz) resolution, respectively, and centered at the redshift determined by \citet{greve12}. A solid red line shows a two-component Gaussian fitting to the CO profiles in each case. The inset plot for SPT 2232-52 shows the comparison of the CO profiles for the eastern and western components.\label{fig:co_line}}
\end{figure*}

\subsection{SPT\,2332-53}

The CO(1-0) line toward SPT\ 2332-53 is well detected and spatially resolved into two main peaks separated by $\sim30''$ (Fig. \ref{fig:co_image}). The CO profiles of both peaks (east and west) are coincident in frequency and line shape suggesting that they arise from the same lensed background object (Fig. \ref{fig:co_line}). A comparison with the SABOCA image from \citet{greve12} indicates that the CO distribution corresponds to the blended IR components A, B, and C, following the nomenclature adopted by Greve et al.

The continuum emission was measured from the 35~GHz correlator window and the line-free channels at $\sim$31.4 GHz. We find significant detections at both frequencies from a source located at $\sim15''$ to the north of the CO emission, with flux densities $S_\mathrm{31.4GHz}=310\pm40\ \mu$Jy and  $S_\mathrm{35GHz}=300\pm35\ \mu$Jy (Fig. \ref{fig:co_image}).This source is identified with the brightest cluster galaxy (BCG) in the foreground structure. It is detected at 3.4 and 4.6 $\mu$m with {\it Spitzer} IRAC, though not at 12 and 22 $\mu$m with the Wide-field Infrared Survey Explorer (WISE). This source is also detected in the 5.5 GHz and 9.0 GHz ATCA images. The non-detection of continuum at the CO position yields a limit to the integrated $\sim35$ GHz flux density of the background source of $<220\ \mu$Jy ($3\sigma$ with respect to the zero level). 

From Gaussian fitting to the CO profile of the combined emission from both components (Fig. \ref{fig:co_line}), we find the fluxes and line-widths listed in Table \ref{table:properties}. The total line intensity is obtained by creating a moment-0 image, integrated over $\sim600$ km s$^{-1}$ around the line (Table $\ref{table:properties}$). 

\subsection{SPT\,0538-50}

The CO(1-0) line toward SPT\ 0538-50 is also significantly detected, however at the spatial resolution of these observations, the source is unresolved (Fig. \ref{fig:co_image}). The spectrum suggests the existence of two CO line peaks (Fig. \ref{fig:co_line}). Fitting a two component Gaussian profile to the spectrum, we find the bright (SPT\ 0538-50blue) and weak (SPT\ 0538-50red) CO components separated by $350\pm50$ km s$^{-1}$ in velocity. The fainter CO component carries $\approx$30\% of the CO intensity. Table \ref{table:properties} summarizes the measurements for these sources. We measure the total CO line intensity from the moment-0 image, integrated over 710 km s$^{-1}$ containing both CO line peaks (SPT\ 0538-50; Table \ref{table:properties}). 

Continuum emission at 30.6 and 35 GHz is also detected (Fig. \ref{fig:co_image}). In this case, the continuum peak position coincides within $1{\farcs}0$ of the location of the CO peak. The origin of this emission is discussed in the next sections. Continuum emission is also detected at 2.2 and 5.5\,GHz, with flux densities of $S_\mathrm{2.2GHz}=650\pm200\ \mu$Jy and $S_\mathrm{5.5GHz}=550\pm100\mu$Jy, respectively.  This source is not detected in our 9.0\,GHz data, resulting in a 3$\sigma$ upper limit of 225\,$\mu$Jy (respect to the zero level).

\section{Analysis}

\subsection{The nature of the SPT sources}

The results of the source reconstruction of SPT\ 2332-53 do not allow us to discern whether this object corresponds to a disk galaxy or merger (Vieira et al in prep.). The CO profile shape, however, does not show evidence for a double-peak structure (Fig. \ref{fig:co_line}) as would be expected for a disk geometry extended on scales of $\sim5-10$ kpc \citep{daddi10} or for an early-stage merging system \citep[e.g., ][]{greve05}. This suggests the molecular gas is distributed over a relatively compact disk or a compact spherical distribution, as in an advanced merger stage, with most of the gas within a radius $\sim1-2$ kpc.

In the case of SPT\ 0538-50, the lens reconstruction shows that this source is composed of two components \citep[see ][]{hezaveh13}: a compact source with a radius of $0.5\pm0.1\,$kpc, and an extended source with a radius of $1.6\pm0.3\,$kpc. The two components appear to spatially overlap, leaving it unclear whether the system corresponds to two interacting galaxies, or a clumpy disk with a bright compact source. However, the fraction of the ALMA 870 $\mu$m flux distributed between both components, $\sim70\%$ of the total $L_\mathrm{IR}$ for the compact source and $\sim30\%$ for the extended one \citep{hezaveh13}, is in good agreement with the CO luminosity budget. The brighter CO component (SPT\ 0538-50blue) carries a luminosity $L_\mathrm{CO}'=1.49\times10^{10}\,(\mu/20)^{-1}$ K km s$^{-1}$ pc$^2$ or $\sim$70\% of the total CO luminosity, while the fainter CO component (SPT\ 0538-50red) has a CO luminosity $L_\mathrm{CO}'=6.1\times10^{9}\ (\mu/20)^{-1}$ K km s$^{-1}$ pc$^2$ or $\sim$30\%. This suggests that the compact reconstructed source could be identified with the bright CO line component, while the extended source could be associated with the fainter, broader CO line component (Fig. \ref{fig:co_line}). However, this IR to CO line association remains uncertain. The only way to confirm this is through spatially resolved CO spectroscopy.

\subsection{Gas and dynamical masses}

The molecular gas mass, $M_\mathrm{gas}$ is commonly computed from the CO(1-0) line luminosity using the relation $M_\mathrm{gas}=X_\mathrm{CO}L_\mathrm{CO}'$, where $X_\mathrm{CO}$ is the CO luminosity to gas mass conversion factor. Its value can vary from galaxy to galaxy, and even within galaxies \citep[e.g.,][]{genzel12}. We adopt $X_\mathrm{CO}=0.8\ M_\odot$ (K km s$^{-1}$ pc$^2)^{-1}$, as found for local ultraluminous infrared galaxies \citep[ULIRGs;][]{downes98}. While this value is typically used for dusty starburst galaxies at high redshift, higher values have been recently found for distant disk galaxies \citep[$X_\mathrm{CO}\sim3.6$;][]{daddi10}. Using a larger value for $X_\mathrm{CO}$ would produce gas masses 4.5 times larger than quoted here. From the CO luminosities found in the previous section, we find $M_\mathrm{gas}=(3.2\pm0.5)\times10^{10} (\mu/15)^{-1} (X_\mathrm{CO}/0.8)\ M_\odot$ and $M_\mathrm{gas}=(1.7\pm0.3)\times10^{10} (\mu/20)^{-1} (X_\mathrm{CO}/0.8)\ M_\odot$ for SPT\,2332-53 and SPT\,0538-50, respectively. 

We compute dynamical masses  for our sources based on the CO line profiles. For a disk gas distribution, the dynamical mass of the system is given by $M_\mathrm{dyn}\approx2.3\times10^5\ \Delta v_\mathrm{FWHM}^2\ R\ \mathrm{sin}^{-2}(i)\ M_\odot$ , where $\Delta v_\mathrm{FWHM}$ is the observed CO line FWHM in units of km s$^{-1}$, $R$ is the disk radius in kpc, and $i$ is the inclination angle. For a merger model, $\Delta v_\mathrm{FWHM}$ represents half the physical separation in velocity of the component CO lines, and $R$ is half the separation between components \citep{solomon05}. 

Given the poor constrains on the geometries of both sources, we derive dynamical masses by adopting disk geometries. For SPT\,2332-53, we use $R=2$ kpc and an average inclination angle $i\sim57\degr$ \citep{law09}, which yields $M_\mathrm{dyn}=7.8\times10^{10}\ M_\odot$. This implies a gas fraction of $\sim0.4$ for $\mu=15$ and the adopted $X_\mathrm{CO}$ value. For SPT\,0538-50, we use $R=1.6$ kpc (from the lens model), a CO profile FWHM of $\sim500$ km s$^{-1}$ (from the full CO line) and an average inclination $i\sim57\degr$, yielding $M_\mathrm{dyn}=1.3\times10^{11}\ M_\odot$. This leads to a gas fraction of $\sim$0.1 for $\mu=20$ and the adopted $X_\mathrm{CO}$. We remark that since the actual source geometries are unknown, the estimated dynamical masses and gas fractions are uncertain and should be taken with caution as they are given for reference only.

\subsection{Star formation efficiencies and depletion timescales}

\begin{figure}
\vspace{3mm}
\centering
\includegraphics[scale=0.5]{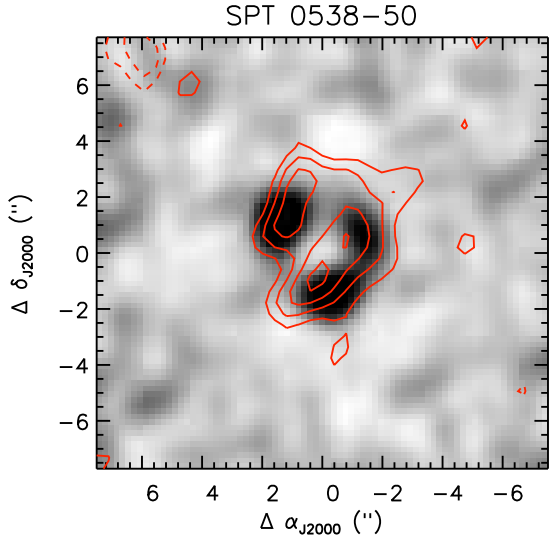}
\caption{The ALMA 345 GHz high-resolution map for SPT\,0538-50 is shown in the background \citep{hezaveh13} with red contours overlaid representing the ATCA 5.5 GHz radio map. Contour levels are in steps of $\pm1\sigma$, starting at $\pm2\sigma$ with $\sigma=35\ \mu$Jy beam$^{-1}$. The spatial coincidence between the submillimeter and radio emission confirms that the latter arises from the background lensed object. \label{fig:atca_alma}}
\end{figure}

The star formation efficiency (SFE) can be defined as SFE $= L_\mathrm{IR}/L_\mathrm{CO}'$ in units of $L_\odot$ (K km s$^{-1}$ pc$^2$)$^{-1}$. From this, we obtain SFEs of 110 and 205 $L_\odot$ (K km s$^{-1}$ pc$^2$)$^{-1}$ for SPT\,2332-53 and SPT\,0538-50, respectively. The gas depletion timescale can be defined as $t_\mathrm{dep}=M_\mathrm{gas}/$SFR. We adopt SFR ($M_\odot$ yr$^{-1}$) $=1.5\times10^{-10}\ L_\mathrm{IR}$ ($L_\odot$) \citep{murphy11} for consistency with \citet{greve12}. This yields gas depletion timescales of $\sim4.9\times10^7 (\mu/15)^{-1} (X_\mathrm{CO}/0.8)$ yr and $\sim2.6\times10^7 (\mu/20)^{-1} (X_\mathrm{CO}/0.8)$ yr for SPT\ 2332-53 and SPT 0538-50, respectively. While values for the depletion timescales for DSFGs and local ULIRGs are typically $<10^8$ yr \citep[e.g.;][]{greve05, solomon05}, distant ``normal'' star forming disk galaxies were found to have depletion timescales of $\sim(0.3-2)\times10^9$ yr \citep{daddi10,tacconi10,tacconi13}. The depletion timescales for our sources appear to be consistent with typical DSFGs and local ULIRGs.

\subsection{SPT\,0538-50: Free-free, synchrotron, and dust emission}

SPT\,0538-50 is additionally detected at 2.2 and 5.5\,GHz by ATCA.  At $z = 2.783$ the 2.2, 5.5, 30.6 and 35\,GHz observations sample 8.3, 20.8, 115.7 and 132.4\,GHz in the rest-frame, respectively, providing a long lever-arm to measure the radio spectral index of the source. 

Fig. \ref{fig:atca_alma} shows the recently obtained ATCA 5.5 GHz continuum image toward SPT\,0538-50.  The image structure matches the position and structure of the Einstein ring found in the ALMA images \citep{hezaveh13}, clear evidence that the bulk of the radio emission comes from the lensed background source. We thus rule out the foreground galaxy contribution to the radio emission based on the morphology shown by ALMA and ATCA.

\begin{figure}
\includegraphics[scale=0.45]{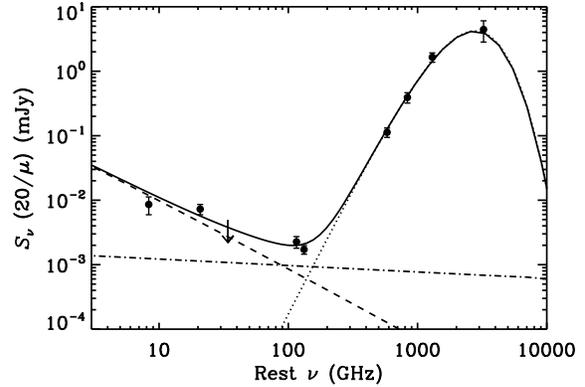}
\caption{The radio-to-IR spectrum of SPT\,0538-50 fit by a combination of modeled radio and IR spectra (solid line). Photometry is shown with error bars, along with an upper limit at $\nu_\mathrm{rest}=$25\,GHz. The IR dust emission is given by a grey-body model (dotted line). The radio spectrum is modeled by 2 components, non-thermal synchrotron emission (long-dashed lines) and thermal, free-free, emission (dot dashed line)\label{fig:radio_sed}}
\end{figure}

In the following section, we analyze the radio spectral properties of SPT\,0538-50, and adopt the rest-frame frequencies for clarity. We define the flux density $S_\nu$ in terms of the observed frequency $\nu$ as $S_\nu \propto \nu^\alpha$, with $\alpha$ the radio spectral index. The data seem to suggest that the radio spectrum may steepen between rest-frame 8.3 and 132.4\,GHz as the observed spectral index between 8.3 and 20.8\,GHz is $-0.18\pm0.39$, and $-0.76\pm0.11$ between 20.8 and 132.4\,GHz (see Fig. \ref{fig:radio_sed}). In fact, subtracting off an estimate of the thermal dust contribution to the 132.4\,GHz flux density suggests an even steeper intrinsic (free-free + synchrotron) radio spectral index between 20.8 and 132.4\,GHz of $-0.89\pm0.14$. 
Given the large uncertainties on the photometry, we do not include an additional free parameter for spectral curvature in our fit to the radio spectrum, and simply assume a single power law without a break.  
However, we consider and discuss the physical implications of the observed spectral flattening at lower frequencies below in Section \ref{sec:spx-flat}.  

\subsubsection{Fitting the radio-to-IR spectrum: free-free emission and an independent SFR estimate}

To estimate the amount of free-free emission at rest-frame $115.7$ and $132.4$\,GHz, which can then be used to calculate a SFR, we fit the full radio-to-IR spectrum. For simplicity, we use the grey-body fit from \citet{greve12} to model the thermal dust component, which has a fixed emissivity spectral index $\beta = 2$. To fit the radio spectrum, we vary a combination of free-free (thermal) and synchrotron (non-thermal) radio emission components which scale as $S_{\nu}^{\rm T} \propto\nu^{\alpha_{\rm T}}$ and $S_{\nu}^{\rm NT} \propto\nu^{\alpha_{\rm NT}}$, respectively, where $\alpha_{\rm T}=-0.1$ and $\alpha_\mathrm{NT}$ are the free-free and synchrotron spectral indices, respectively. 

Before varying the free-free and synchrotron components to fit the observed radio spectrum, we first need to assume a non-thermal spectral index.  We assume a synchrotron spectral index of $\alpha_{\rm NT}=-1.0$, which is consistent with the steepest part of the radio spectrum measured between 20.8 and 132.4\,GHz (see above). This value is in agreement with an estimate for the non-thermal spectrum based on the calculation of energy losses to cosmic ray electrons in this system arising from synchrotron radiation, inverse Compton scattering, ionization, bremsstrahlung, and escape through an empirical prescription \citep[see ][for details]{murphy09b}. Furthermore, this value is consistent with the synchrotron spectral index of $-1.05\pm0.05$ for two dusty star-forming galaxies at $z\sim2.5$ found by \citet{thomson12}. Setting $\alpha_{\rm NT}=-1.0$, our best fit model to the radio-to-IR data suggests a free-free radio fraction at 115.7 and 132.4\,GHz of $42\pm10$\% and $55\pm13$\%, respectively (see Fig. \ref{fig:radio_sed}). 

Taking these values and the free-free SFR calibration given in \citet{murphy12} results in a ${\rm SFR}=470\pm170\,M_{\odot}\,{\rm yr}^{-1}$. This is consistent with the IR luminosity-derived estimate of ${\rm SFR}=650\pm100\,M_{\odot}$ yr$^{-1}$, which is again calculated using the calibration given in \citet{murphy11}, calibrated for a common IMF \citep[i.e., ][]{kroupa01}, and consistent with the estimate from \citet{greve12}. We note that such calibrations between IR luminosity and SFR are not likely accurate to better than a factor or 2 \citep[][]{kennicutt98,omont01}, and this uncertainty is not included in the error above on the IR-derived SFR. 

\subsubsection{Physical considerations in the fitting}

While we have assumed a physically motivated value for the non-thermal spectral index, it is worth noting that this assumed value is marginally inconsistent with a naive measurement of $\alpha$ from our data. Using an ordinary-least-squares fit to the four radio detections, weighted by the uncertainties of the flux density measurements, results in a radio spectral index (before correcting for thermal dust emission) of  $\alpha=-0.65\pm0.10$. Assuming this value for the non-thermal radio spectral index, the fit  results in a 132.4\,GHz free-free fraction of $\sim$2\%. Similarly, using a dust emissivity index $\beta=2$ and a radio spectral index $\sim0.75$ \citep{ibar10} would result in a free-free fraction of $\sim$20\%.

For any physical configuration in which most of the IR emission is powered by star formation, such a small fraction of free-free radio emission is highly unlikely. A situation without free-free emission could arise for cases where: (a) the emission is purely powered by an AGN; (b) there is a suppression of Lyman continuum photons, and thus free-free emission, through an upper mass cutoff of the IMF at $\sim20\ M_\odot$ while still providing synchrotron emission from supernovae ($\sim8-20\ M_\odot$); and (c) the absorption of ionizing photons by dust in the actual starburst. However, the galaxy shows a rather normal, albeit slightly higher, IR/radio ratio (see Section \ref{sec:spx-flat}), which is difficult to keep constant if there are variations in the IMF or in the presence of a powerful AGN. In the case where dust successfully competes with neutral hydrogen for ionizing photons, we still expect a significant free-free component to be present and compatible with the amount of ongoing star formation inferred from the IR spectrum \citep{murphy12}. A similar situation has been observed in local (U)LIRGs, where the synchrotron spectrum is found to be increasingly depressed at high frequencies \citep[e.g.,][]{clemens08, leroy12}, perhaps arising from a modified electron injection spectrum in such dense starbursts.  

\subsubsection{Spectral Flattening Due to a Deeply Embedded Compact Starburst \label{sec:spx-flat}}

The radio observations suggest that the radio spectrum flattens towards lower frequencies.  This flattening may be the result of the radio continuum emission becoming optically thick at lower frequencies, as observed in local compact ULIRGs \citep{condon91}. Indeed, the lens model of SPT\,0538-50 indicates the presence of an IR-dominant, compact source. Such a spectral flattening at lower radio frequencies typically causes sources to have IR/radio (1.4\,GHz) ratios that are larger than the average value among star-forming galaxies \citep{condon91}. Extrapolating a 1.4\,GHz flux density using the observed rest-frame 8.3\,GHz flux density along with the spectral index measured between 8.3 and 20.8\,GHz suggests that the galaxy has a logarithmic IR/1.4\,GHz ratio $q_{\rm IR} = 2.84$\footnote{The latter is defined as the ratio of the total IR luminosity ($8-1000\mu$m) to the radio power: 
\begin{align*}
q_{\rm IR}={\rm log}_{10} \left( \frac{L_\mathrm{IR}}{3.75\times10^{12} \mathrm{W~m}^{-2}} \right) - {\rm log}_{10}\left( \frac{ S_{\rm 1.4 GHz}} {{\rm W~m}^{-2}~{\rm Hz}^{-1}} \right). \nonumber
\end{align*}}.
This implies a linear ratio nearly 60\% (1$\sigma$) larger than the average value of $q_{\rm IR}=2.64\pm0.02$ found for local star-forming galaxies with 1.4 GHz luminosities in the range $\sim 10^{19} - 10^{24.5}$ W~Hz$^{-1}$\citep{bell03}.

Assuming that the spectral flattening is in fact the result of the source being powered by an optically-thick starburst, we can use local relations to infer properties of the galaxy. Recently \citet{murphy13} found a correlation between the radio spectral index of $z\sim0$ (U)LIRGs and their distance from the star-forming main sequence (i.e., specific SFR defined as $\mathrm{sSFR} = \mathrm{SFR}/M_{\star}$). That is, galaxies having flatter spectral indices, most likely due to harboring compact starbursts, which are optically thick at low radio frequencies, have increasingly higher sSFRs.  Based on this correlation, and assuming an evolution in the sSFR of galaxies from $z\sim0$ to $z\sim3$, where the typical sSFR at $z\sim3$ is 4.5\,Gyr$^{-1}$ \citep{karim11}, the spectral index measured between 8.3 and 20.8\,GHz suggests that the sSFR of SPT\,0538-50 is $\sim 10.2\pm2.9$ Gyr$^{-1}$, more than a factor of $\sim$2 above the $z\sim3$ main sequence for star-forming galaxies \citep{karim11}. Taking the above free-free estimate for the SFR suggests this system has $M_\mathrm{\star}= (4.6\pm2.1)\times 10^{10}\,M_{\odot}$.

\section{Discussion and Summary}

We have obtained sensitive CO(1-0) and radio continuum ATCA observations toward two bright DSFGs discovered in millimeter survey data taken with the SPT. This corresponds to a pilot study of a larger sample of millimeter bright SPT sources with available spectroscopic redshifts.

Both systems are found to be rich in molecular gas with masses $\sim2\times10^{10} M_\odot$, gas fractions of $\sim0.1-0.4$ and SFEs of $\sim100-200$ $L_\odot$ (K km s$^{-1}$ pc$^2$)$^{-1}$, for the adopted values of the magnification ($\mu$) and CO luminosity to gas mass conversion factor ($X_\mathrm{CO}$). Taken together, the observed CO properties are compatible with those found in other lensed (and unlensed) submillimeter-selected galaxies based on previous CO(1-0) measurements  \citep{frayer11, ivison11, harris10}. 

In both cases, we find $\sim31$ and 35 GHz continuum detections in the field (rest-frame 115.7 and 132.4\,GHz). While in the case of SPT\,2332-53 the continuum emission comes from the foreground BCG galaxy, in the case of SPT\,0538-50 the radio emission is clearly identified with the CO source, as shown in Fig. \ref{fig:atca_alma}. Assuming a non-thermal spectral index of $\alpha_\mathrm{NT}=-1.0$, our modeling of the radio to IR spectrum of the SPT\,0538-50 suggests that $42\pm10\%$ and $55\pm13\%$ of the emission at rest-frame $115.7$ and 132.4 GHz, arise from a free-free component. This allows us to derive a free-free SFR value of $470\pm170~M_\odot$ yr$^{-1}$, which is consistent within the uncertainties with the IR-derived SFR. 
Recently, \citet{thomson12} detected CO(1-0) line emission in three submillimeter-selected galaxies at $z\sim2.5-2.9$, and marginally detected the rest-frame $115$ GHz continuum from two of these galaxies. Their modeling of the radio spectra indicates free-free contributions to the rest-frame 115 GHz in the range $\sim30-40\%$ \citep{thomson12}. This is similar to our results, which are based on significant detections at rest-frame 115.7 and 132.4 GHz on SPT\,0538-50.

Analysis of the radio spectral curvature for this source suggests that the star formation activity is powered by a compact starburst; the spectral flattening towards lower radio frequencies is similar to what is observed for local (U)LIRGs harboring compact starbursts that become optically-thick at low radio frequencies. From this analysis, we estimate a stellar mass of $(4.6\pm2.1)\times10^{10}\,M_{\odot}$. It is interesting to note that almost all local ULIRGs are powered by merger-driven star formation and lie above the star-forming main sequence, which also seems to be the case for SPT\,0538-50 with a sSFR of 10.2 Gyr$^{-1}$ compared to the main-sequence at $z\sim3$ with an average sSFR of 4.5 Gyr$^{-1}$. Furthermore, the ULIRG-like nature of this source is supported by the results from the lens model which indicate that SPT\,0538-50 is composed of a compact IR-dominant component, which comprises 70\% of the IR luminosity \citep{hezaveh13}. This is similar to what is seen in local ULIRGs \citep{bryant99}.

These results exemplify the use of deep low-J CO spectroscopy in high-redshift star forming galaxies taking advantage of the large bandwidths available with ATCA and the Karl Jansky Very Large Array (VLA). The advent of important samples of bright, gravitationally lensed galaxies from the SPT and {\it Herschel} surveys, make possible detailed studies of the ISM in DSFGs. While observation of the CO line emission enables a direct measurement of the gas and dynamical masses, modeling of the radio spectrum enables independent measurements of the SFRs and stellar masses without the need of near-to-far IR imaging, which generally requires space-based observatories.

\section*{Acknowledgments}

We thank Andy Biggs for useful discussions. MA and this work was co-funded under the Marie Curie Actions of the European Commission (FP7-COFUND). The Australia Telescope is funded by the Commonwealth of Australia for operation as a National Facility managed by CSIRO. The SPT is supported by the National Science Foundation through grant ANT-0638937, with partial support through PHY-1125897, the Kavli Foundation and the Gordon and Betty Moore Foundation.

\bibliographystyle{mn2e}
\bibliography{../../bibtex/spt_smg}

\bsp

\label{lastpage}

\end{document}